\documentclass[%
 reprint,
preprintnumbers,
 amsmath,amssymb,
 aps,
]{revtex4-1}

\usepackage{graphicx}
\usepackage{dcolumn}
\usepackage{bm}
\usepackage[dvipsnames,usenames]{xcolor}
\usepackage{amsmath}
\usepackage{amssymb} 
\usepackage{hyperref}
\usepackage{bbold}
\usepackage{algpseudocode}

\input{Qcircuit}

\begin{document}

\preprint{LA-UR-18-22120}

\title{Linear Response on a Quantum Computer}

\author{Alessandro Roggero}
 \email{roggero@lanl.gov}
\author{Joseph Carlson}
 \email{carlson@lanl.gov}
\affiliation{
Theoretical Division, Los Alamos National Laboratory, Los Alamos, New Mexico 87545, USA
}%

\date{\today}

\begin{abstract}
The dynamic linear response of a quantum system is critical for understanding both the structure and dynamics of
strongly-interacting quantum systems, including neutron scattering from materials, photon and electron scattering
from atomic systems, and electron and neutrino scattering by nuclei. We present a general algorithm for quantum 
computers to calculate the dynamic linear response function with controlled errors and to obtain 
information about specific final states that can be directly compared to experimental observations.
\end{abstract}


\maketitle


Quantum computers should enable dramatic new capabilities in simulating quantum many-body systems, particularly their dynamic properties~\cite{Feynman1982}.
Quantum dynamics is in general extremely difficult to treat on a classical computer except for a few special cases such as very low-energy scattering 
where spectral decomposition in finite volumes enable direct connections between spectra and phase shifts
in the scattering of 2- or 3-clusters~\cite{Luscher1986,Luscher1991, Carlson1987, Sharpe2015} or very high-energy 
scattering that can be treated 
as nearly non-interacting final states, including y-scaling in neutron
or electron scattering~\cite{West1975,Sears1984} or inclusive deep inelastic scattering in QCD~\cite{dokshitzer1977}. The general problem is essentially intractable because of quantum interference,
the rapidly oscillating phases that arise in the relevant path integrals.

Perhaps the simplest quantum dynamics problem is the dynamic linear response, framed as the response of a quantum system to a small perturbation. Examples are ubiquitous,
including for example neutron scattering on materials, photon scattering in atomic systems, and electron and neutrino scattering from atomic nuclei. The response
of the system can in principle tell us much about the structure of the system being probed as well as important properties of the dynamics.  In the case of
neutrinos scattered by nuclei it is also used to infer properties of the neutrino itself  including 
masses, mixing angles, the mass hierarchy and CP violation in the neutrino sector~(eg. \cite{refdune}).

The ability to accurately calculate the dynamic response over a wide range of energy and momentum transfers, augmented by the possibility of determining specific
features of the final states, would revolutionize our ability to extract information from many kinds of scattering experiments.
Some information on quantum dynamics can be obtained using classical computers even for relatively large systems, typically by computing imaginary-time
correlation functions~\cite{Gubernatis1991, Carlson1992, Ceperley1995}. Even for systems where the ground state or thermal ensembles can be simulated free of any
sign problem, it is extremely difficult to invert these correlation functions to obtain the exact dynamic response. In this paper we discuss methods to determine
the dynamic response on a quantum computer, as well as to detect important features of explicit final states that can be directly compared to experimental data.
Our approach is particularly well suited for problems defined on a lattice, but these lattice methods can of course also be used to simulate systems in the continuum over a wide
but finite range of energy and momenta, for example in lattice studies of cold atoms~\cite{Bulgac2006, Carlson2011} and nuclear systems~\cite{Lee2004,Lee2009}.  Also we restrict ourselves to the response from the quantum ground state (T=0), generalizations
to finite temperature are possible by preparing states in thermal equilibrium rather than the ground state~\cite{Temme2011}.

We note that the similar problem of evaluating chemical reaction rates \cite{Lidar1999,Kassal2008} and time-dependent correlation functions \cite{Terhal2000,Ortiz2001,Somma2002} have been already investigated in the quantum-chemistry literature. Our proposed algorithm improves on these earlier techniques in that our strategy is completely general, does not depend on simplifying assumptions on the excitation operator (for example, being able to diagonalize it as in \cite{Terhal2000}) and requires only a polynomial number of measurements (instead of exponential like in \cite{Lidar1999}). Also, working directly in frequency space allows us a direct access to the final states of a reaction which can be further analyzed. This is particularly important for neutrinos where the momentum and energy transfer are {\it a priori} unknown.

Furthermore we are able to provide rigorous cost and error estimates of the computed dynamical properties.
Available algorithms for evaluating energy spectra \cite{Somma2002,Wang2012} can in principle be adapted to compute response functions but they require resolution of individual excited states which grows exponentially in number for large systems.

The paper is organized as follows, in Sec.~\ref{sec:method} we provide detailed definition of the Dynamical Response Function and describe the implementation of our method.
In Sec.~\ref{sec:obs} we provide an example of final state characterization by discussing the estimation of the one- and two-body momentum distribution and conclude in Sec.~\ref{sec:conclusions}. 

\section{\label{sec:method}Method}
In the linear regime the response of a system of interacting particles due to a perturbative probe characterized by the excitation operator $\hat{O}$ can be fully characterized using the Dynamical Response Function, which can be expressed as
\begin{equation}
S_{O}(\omega) = \sum_\nu \lvert\langle \psi_\nu\lvert\hat{O}\rvert\psi_0\rangle\rvert^2\delta(E_\nu-E_0-\omega)
\end{equation}
where $\rvert\psi_0\rangle$ is the ground-state of the system with energy $E_0$, $\rvert\psi_\nu\rangle$ are the final states of the reaction with energies $E_\nu$ and $\omega$ is the energy transfer. It is convenient to rescale the response function so that it's zero moment (the integral over frequencies) is $1$; this can be achieved by defining
\begin{equation}
\label{eq:scaled_resp}
S^r_{O}(\omega) = \sum_\nu \frac{\lvert\langle \psi_\nu\lvert\hat{O}\rvert\psi_0\rangle\rvert^2}{\langle\hat{O}^2\rangle_0}\delta(E_\nu-E_0-\omega)\;.
\end{equation}
The final normalization can be restored by either using the knowledge of one of the sum rules or by direct evaluation of the ground state expectation value $\langle\hat{O}^2\rangle_0\equiv\langle\psi_0\lvert\hat{O}^2\rvert\psi_0\rangle$. Understanding this, in the following we will drop the superscript $r$. 

Our goal is to estimate the dynamical response function $S_O(\omega)$ with energy resolution $\Delta\omega$ and a precision $\delta_S$ with probability $1-\epsilon$. We will indicate the difference between the largest eigenvalue of $\hat{H}$ and the ground state energy by: $\Delta H=E_{max}-E_0$. Note that this quantity grows only polynomially with system size for most Hamiltonians of interest (see discussion below).

In the following we will assume to have access to three black-box quantum procedures (oracles):
\begin{itemize}
\item a unitary $\hat{U}_G$ which prepares the ground-state of the Hamiltonian of interest
\item a unitary $\hat{U}_O$ which implements time evolution under $\hat{O}$ for a short time $\gamma < poly(\delta_{S})$
\item a unitary $\hat{U}_t$ which implements time evolution under the system Hamiltonian for time $t$
\end{itemize}

Even though the oracle $\hat{U}_G$ may be impractical to implement for a general Hamiltonian, for most systems of interest many different algorithms are available in the literature (\cite{Farhi2000,Aspuru-Guzik2005,Poulin2009,Ward2009,Peruzzo2014,Shen2017,Hefeng2016,Kaplan2017,Berry2017}) and some have already be tested on simple nuclear systems~\cite{Dumitrescu2018}. Also, close to optimal strategies to implement the time-evolution operator for sparse Hamiltonians are known \cite{Berry2015a,Berry2015b} and for Hubbard-type Hamiltonians (like those derived within lattice-EFT~\cite{Lee2009}) efficient implementations of Trotter steps with sub-linear circuit depth are available \cite{Kivlichan2017}. For the common case where $\hat{O}$ is a one-body operator the latter strategies can be used to implement $\hat{U}_O$ efficiently.

Our scheme is composed of two quantum circuits
\begin{itemize}
\item a state preparation routine requiring $\mathcal{O}(1)$ calls to $\hat{U}_G$ and $\hat{U}_O$ with a success probability (see Sec.~\ref{sec:state_prep})
\begin{equation}
\label{eq:psuccess}
P_{success}=\mathcal{O}\left(\delta_S\frac{\langle\hat{O}^2\rangle_0}{\|\hat{O}\|^2}\right)
\end{equation}
where $\langle\cdot\rangle_0$ denotes the expectation value on the ground state and $\|\cdot\|$ is the operator norm;
\item a second routine that provides access to $S_O(\omega)$ which requires $W=log_2\left(\Delta\omega/\Delta H\right)$ ancilla qubits, the application of $\hat{U}_t$ for a maximum time $t_{max}=2\pi/\Delta\omega$ and additional $\mathcal{O}\left(Wlog(W))\right)$ gates
\end{itemize}

For typical situations where the implementation of $\hat{U}_G$ requires considerable effort the success probability of the first routine can be amplified to $\mathcal{O}(1)$ with additional $\mathcal{O}(1/P_{success}^2)$ calls to the oracle $\hat{U}_O$. An alternative algorithm which removes the dependence of $P_{success}$ on $\delta_S$ but is more difficult to make deterministic is also presented in Sec.~\ref{sec:state_prep}.

This whole circuit needs to be run a number of times given approximately by
\begin{equation}
\label{eq:nrep}
N_{rep}\approx ln\left(\frac{2}{\epsilon}\right)\frac{1}{2\delta_S^{2}}
\end{equation}
independent of the target resolution $\Delta\omega$.

In summary, for a given choice of the excitation operator $\hat{O}$ our algorithm can be described by the following steps:
\begin{center}
\begin{algorithmic}
\While {iteration number less than $N_{iter}$}
  \State prepare the ground state using $\hat{U}_G$\;
  \State run the first quantum algorithm (Sec.~\ref{sec:state_prep})\;
  \If {algorithm succeeds}
    \State we have prepared $\rvert \Phi_O \rangle \propto \hat{O}\rvert \psi_0 \rangle$\;
    \State run the second quantum algorithm (Sec.~\ref{sec:resp})\;
    \State store result for classical post-processing\;
    \If {final state information needed}
      \State measure final state (eg. Sec~\ref{sec:obs})\;
    \EndIf
  \EndIf
\EndWhile
\end{algorithmic}
\end{center}
In the next sections we describe in detail the implementation of the two quantum routines introduced above. We also present examples obtained by classical simulation of a simple 2D fermionic system described by the Hubbard hamiltonian
\begin{equation}
\label{eq:hubbard}
\begin{split}
H&= -t\sum_{\sigma=1}^2\sum_{\langle i,j\rangle}^M\left( c^\dagger_{i,\sigma}c_{j,\sigma}+c_{i,\sigma}c^\dagger_{j,\sigma}\right)\\
& +U \sum_{i=1}^M \hat{n}_{i,\uparrow}\hat{n}_{i,\downarrow}\;,
\end{split}
\end{equation}
where $\langle i,j\rangle$ indicates the nearest-neighbor lattice sites and $\hat{n}_{i,\sigma}=c^\dagger_{i,\sigma}c_{i,\sigma}$ denotes the number operator.
The results shown here were obtained for $A=2$ "nucleons", $M=31^2$ lattice sites and $U/t=-2$. These parameters are chosen to give a bound state considerably smaller than the lattice.

\subsection{\label{sec:state_prep}State preparation algorithm}
The first problem we have to solve is the preparation of the state $\rvert \Phi_O\rangle$ given a quantum register initialized in the ground-state $\rvert \psi_0\rangle$. Let's start by adding an ancilla qubit and defining the unitary operator
\begin{equation}
\label{eq:oprop}
\hat{U}^\gamma_S =e^{-i\gamma \hat{O}\otimes\sigma_y}=\begin{pmatrix}
cos(\gamma\hat{O})&-sin(\gamma\hat{O})\\
sin(\gamma\hat{O})&cos(\gamma\hat{O})\\
\end{pmatrix}
\end{equation}
where the Pauli $\sigma_x$ operator acts on the ancilla and the final matrix representation is on the basis spanned by the states $\{\rvert0\rangle,\rvert1\rangle\}$ of the ancilla. Note that this unitary can be implemented efficiently with just $2$ calls to a controlled version of the oracle $\hat{U}_O$ and additional $\mathcal{O}(1)$ one-qubit gates. 

By initializing the ancilla register to $\rvert1\rangle$, applying $\hat{U}^\gamma_S$ and measuring the state $\rvert0\rangle$ we have effectively produced
\begin{equation}
\label{eq:state_prep}
\left(\mathbb{1}\otimes\rvert0\rangle\langle0\lvert\right)\hat{U}^\gamma_S\rvert\psi_0\rangle\otimes\rvert1\rangle=\!\frac{\rvert\Phi_O\rangle}{\sqrt{\langle\Phi_O\vert\Phi_O\rangle}}+\mathcal{O}\left(\gamma^2\|\hat{O}\|^2\right)
\end{equation}
which differs from the wanted state by corrections of order $\gamma^2$. The error in the implementation of the unitary $\hat{U}_O$ needs to be at least of the same order, which means a simple single Trotter step will suffice.
The state preparation has a success probability of
\begin{equation}
\begin{split}
\label{eq:pofzero}
P_{success}=P(\rvert0\rangle) &= \langle\psi_0\lvert sin(\gamma\hat{O})^2\rvert\psi_0\rangle \\
&= \gamma^2 \langle\hat{O}^2\rangle_0 + \mathcal{O}\left(\gamma^4\right)\;.
\end{split}
\end{equation}

This approach for the application of a non-unitary transformation is similar in spirit to earlier work (see eg. \cite{Williams2004,Terashima2005}) and it suffers from a possibly very low efficiency since we may need $\mathcal{O}(1/\gamma^2)$ trials to succeed. One option is to perform the algorithm at a few relatively large values of $\gamma$ and fit a quadratic function to extrapolate out the error from the final response function. This approach is however complicated if one is interested also in properties of the final states. A second approach, already proposed in \cite{Williams2004}, is to repeat the application of the unitary $\hat{U}^\gamma_S$ until success. This works because $cos(\gamma\hat{O})$ is approximately the identity. In order to obtain a success probability $P(\rvert0\rangle)=\mathcal{O}(1)$ we will need $\mathcal{O}(1/\gamma^2)$ repetitions. In addition, if the inverse $\hat{O}_G^\dagger$ of the ground-state preparation circuit is available then it's possible to use Amplitude Amplification~\cite{Brassard2002} to gain a quadratic speedup over this \footnote{Note that one cannot apply Oblivious Amplitude Amplification \cite{Berry2015a,Berry2015b} since $sin(\gamma\hat{O})$ is not unitary.}.

Note that by using the normalized state $\rvert\Phi_O^\gamma\rangle$ we will compute the normalized response function Eq.~\eqref{eq:scaled_resp}. If no sum-rules are known one can estimate the normalization constant by estimating the success probability Eq.~\eqref{eq:pofzero} at different values of $\gamma$ and extrapolating.

Since the state preparation through the unitary $\hat{U}^\gamma_S$ is only approximate, the parameter $\gamma$ would need to depend on the final target accuracy. As mentioned in the introduction an alternative scheme that avoids this problem by removing the error in Eq.~\eqref{eq:state_prep} can be obtained by representing the excitation operator $\hat{O}$ as a linear combination of $D$ unitary matrices
\begin{equation}
\hat{O}=\sum_{k=1}^D \alpha_k \hat{U}_k\quad\alpha=\sum_{k=1}^D \lvert\alpha_k\rvert \geq \|\hat{O}\|
\end{equation}
which can be efficiently implemented employing additional $m=log_2(D)$ ancilla qubits using known techniques \cite{Berry2015a,Berry2015b,Hao2016}. The success probability in this case is given by
\begin{equation}
\bar{P}_{success}=\frac{\langle\hat{O}^2\rangle_0}{\alpha^2}
\end{equation}
which depending on the particular case may be larger than Eq.\eqref{eq:psuccess}. The main drawback of this approach is that Amplitude Amplification is the only process that can make the algorithm deterministic since upon failure the output state can in general be very different from the starting point. 

\subsection{\label{sec:resp}Response Function estimation}

We now present our strategy to obtain the response function trough the standard Phase Estimation Algorithm (PEA)~\cite{Abrams1999}. It is convenient to shift and scale the original Hamiltonian:
\begin{equation}
\overline{H}=\frac{H-E_0}{\Delta H}\;\Rightarrow\;\overline{H}\rvert\psi_\nu\rangle=\lambda_\nu\rvert\psi_\nu\rangle
\end{equation}
so that we map the energy spectrum to $\lambda_\nu\in[0,1]$.


By direct calculation we see that the response function $\overline{S}_O(\overline{\omega})$ obtained from $\widetilde{H}$ is related to the original one by
\begin{equation}
\label{eq:resp_rel}
\Delta H S_O(\omega)=\overline{S}_O(\overline{\omega})\;,
\end{equation}
for a scaled frequency $\overline{\omega}\in[0,1]$.

Our goal is to estimate $\overline{S}_O(\overline{\omega})$ efficiently. We do this by using PEA on an auxiliary register of $W$ qubits with the evolution operators
\begin{equation}
U^k=e^{i2k\pi\widetilde{H}}\;\Rightarrow\;U^k\rvert\psi_\nu\rangle=e^{i2k\pi\lambda_\nu}\rvert\psi_\nu\rangle
\end{equation}
for $k=0,\dots,2^W-1$. The resulting circuit will have depth $\mathcal{O}\left(Wlog(W)+N_{t_{max}}\right)$, where the first term comes from the inverse Quantum Fourier Transform~\cite{Hales2000} and  $N_{t_{max}}$ is the gate count needed for a time evolution of $t_{max}=\mathcal{O}\left(2\pi/\Delta\omega\right)$ using the oracle $\hat{U}_t$.
The resulting probability of measuring the $W$ ancilla qubits in the binary representation of the integer $y\in[0,2^W-1]$ is (see eg.~\cite{Cleve1998} for more details)
\begin{equation}
\begin{split}
P(y)&=\frac{1}{2^{2W}}\sum_{\nu}\lvert \langle \psi_\nu\vert\Phi_O\rangle\rvert^2 \frac{sin^2\left(2^W\pi\left(\lambda_\nu-\frac{y}{2^W}\right)\right)}{sin^2\left(\pi\left(\lambda_\nu-\frac{y}{2^W}\right)\right)}\\
&\equiv \frac{1}{2^{W}} \sum_{\nu}\lvert \langle \psi_\nu\vert\Phi_O\rangle\rvert^2 F_{2^W}\left(2\pi\left(\lambda_\nu-\frac{y}{2^W}\right)\right)
\end{split}
\end{equation}
where $F_{2^W}(x)$ is the well-known Fejer kernel from Fourier analysis (see eg.~\cite{jerri1998gibbs}). The probability distribution $P(y)$ is a good approximation of $\overline{S}_O(\overline{\omega})$ since this kernel can be seen as a representation of the delta function with width $\Delta x\approx 2^{-W}$. Therefore if we require a frequency resolution $\Delta\omega$ we will need $W=log_2\left(\Delta H/\Delta \omega\right)$ auxiliary qubits and a polynomial number of applications of the time evolution operator to obtain a sample from $P(y)$.

As mentioned above, for most Hamiltonians of interest the energy gap $\Delta H$ scales only  polynomially with the size of the system. 

We now need to estimate $P(y)$ from $N$ samples drawn from it. Since $y$ is a discrete variable an efficient way of reconstructing the probability distribution is by producing an histogram $h_N(y)$ from the samples. Using Hoeffding's inequality~\cite{Hoeffding63} we find that
\begin{equation}
Pr\left(\lvert h_N(y) - P(y)\rvert\geq\delta\right)\leq 2e^{-2N\delta^2}\;,
\end{equation}
which implies in order to obtain a precision $\delta$ with probability $1-\epsilon$ we need approximately
\begin{equation}
N=ln\left(\frac{2}{\epsilon}\right)\frac{1}{2\delta^2}
\end{equation}
independent samples.

\begin{figure}[htbp] 
\centering
\includegraphics[width=0.9\linewidth]{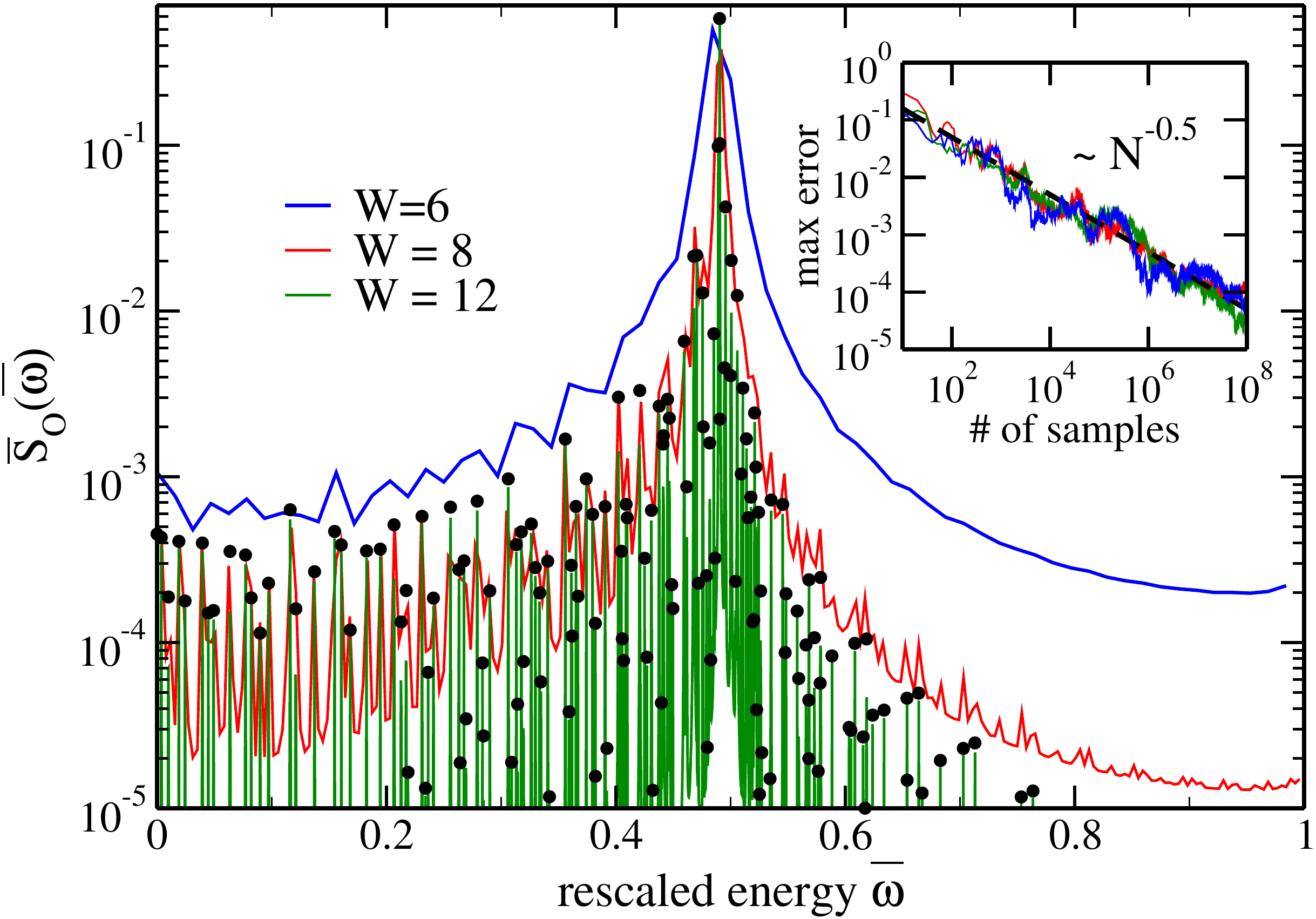} 
\caption{Approximations of the true response function $\overline{S}_O(\overline{\omega})$ for the model system described by the hamiltonian of Eq.~\eqref{eq:hubbard} for different numbers of the work qubits: $W=6$ (blue line), $W=8$ (red line) and $W=12$ (green line). The exact response is also shown with black dots. The inset shows the maximum error in the sample estimate of $P(y)$ as a function of the number of samples.}
\label{fig:response}
\end{figure}

In Fig.~\ref{fig:response} we plot the approximate response $P(y)$ for the model Hamiltonian Eq.~\eqref{eq:hubbard} at three different values of $W$ (6,8,12). By comparing with the exact result shown as black dots, we see that for $W=12$ the effect of energy resolution is negligible but already with $W=8$ we obtain a rather accurate estimate for $\overline{S}_O(\overline{\omega})$. Even $W=6$ reproduces important features of the response, which in experiments is convoluted with the detector resolution. The inset shows the convergence of the maximum error
\begin{equation}
\delta_{max}=\sup_{y\in[0,\dots,2^W-1]}\lvert h_N(y) - P(y)\rvert
\end{equation}
as a function of the sample size $N$. Response functions relevant for $\nu$ and $e^-$ scattering are typically smooth at high energy and hence require small $W$ and short propagation times.

Finally, in order to obtain a negligible bias from the state preparation we need the parameter $\gamma$ to scale as
\begin{equation}
\gamma\lessapprox C\frac{\sqrt{\delta}}{\|\hat{O}\|}
\end{equation}
for some constant $C=\mathcal{O}(1)$. Note that the Hamiltonian evolution implemented in $\hat{U}_t$ has to have an error $\epsilon_t\leq \gamma^2\|\hat{O}\|^2$ to be negligible (luckily algorithms with only logarithmic dependence on $\epsilon_t$ are known \cite{Berry2015a,Hao2016}).

This concludes the proof of the scalings \eqref{eq:psuccess} and \eqref{eq:nrep}.
\section{\label{sec:obs}Final state measurements}
In electron- or neutrino-nuclear scattering experiments~\cite{Benhar2008,Subedi2008,Korover2014,Hen2014,refjlab,refnova,reficarus,refmicroboone,refminerva,refsbnd,refdune,refminosplus,refminos,refargoneut,refnutev} one would like to infer the probability $P(q,\omega\vert\vec{p})$ that the probe transferred energy-momentum $(q,\omega)$ to the nucleus and simultaneously that the final state includes a nucleon (or neutron or proton) of momentum $(\vec{p})$. More concretely this amounts to an inference procedure of the form
\begin{equation}
\begin{split}
P(q,\omega\vert\vec{p})&=P(\vec{p}\vert q,\omega)\frac{P(q,\omega)}{P(\vec{p})}\\
&=P(\vec{p}\vert q,\omega)\frac{P(\omega\vert q)P(q)}{P(\vec{p})}
\end{split}
\end{equation}
where $P(\vec{p})$ results from the experimental measure, $P(\vec{p}\vert q,\omega)$ is the momentum distribution of the final states for a process with given $(q,\omega)$ and $P(q\vert\omega)\equiv S(q,\omega)$. The prior probability $P(q)$ depends on the static response of the nucleus and the characteristic of the probe beam and can be updated given the other ones by a Bayesian procedure. The above section explains how to obtain $S(q,\omega)$ with a given accuracy and in the following we will show how to evaluate few-body momentum distributions given by the final state of the algorithm above. Note that after measuring the $W$ ancilla qubits of Sec.\ref{sec:resp} the main register will be left in a state $\rvert\Psi_f\rangle$ composed by a linear superposition of final states corresponding to energy transfer $\omega\pm\Delta\omega$. Imagine we want now to compute exclusive 1 and 2-body momentum distributions
\begin{equation}
n_1(A)=\langle\Psi_f\lvert\hat{n}_A\rvert\Psi_f\rangle\quad n_2(A,B)=\langle\Psi_f\lvert\hat{n}_A\hat{n}_B\rvert\Psi_f\rangle
\end{equation}
where $\hat{n}_k\equiv\hat{n}(\vec{p}_k,\sigma_k,\tau_k)$ is the number operator for a state with momentum $\vec{p}_k$, spin $\sigma_k$ and isospin $\tau_k$. We can define a unitary operator $U_{n_A}=exp(-i\pi\hat{n}_A)$ (which is efficiently implementable) and run the following circuit with an ancilla qubit
\begin{equation}
\Qcircuit @C=1em @R=.7em {
\lstick{\ket{0}}&\gate{H} & \ctrl{1}  & \gate{H} &\qw &\meter\\
\lstick{\ket{\Psi_f}}& \qw&\gate{U_{n_A}} &  \qw &\qw&\qw
}
\label{circuit:mom_dis}
\end{equation}
By using the idempotence of $\hat{n}_A$ we find
\begin{equation}
P(\rvert0\rangle)=1-n_1(A)\quad P(\rvert1\rangle)=n_1(A)
\end{equation}
and we can then extract the expectation value by estimating these probabilities. Note that we may use the same procedure with $U_{n_A,n_B}=exp(-i\pi\hat{n}_A\hat{n}_B)$ to estimate $n_2(A,B)$ (and possibly higher body momentum distributions). We can get a better strategy by reusing the final state of circuit \eqref{circuit:mom_dis} upon measuring the ancilla in $\rvert1\rangle$ and running it again with $U_{n_B}$ since the probabilities now will be
\begin{equation}
P'(\rvert0\rangle)=1-\frac{n_2(A)}{n_1(A)}\quad P'(\rvert1\rangle)=\frac{n_2(A)}{n_1(A)}\;.
\end{equation}

Note that $\rvert\Psi_f\rangle$ will in general be contaminated by final state interactions but we can access a better approximation to an asymptotic state by evolving it in time using $\hat{U}_t$.

This measurement procedure will need to then be repeated a polynomial number of times for all the observables of interest. Given the expensive procedure needed to generate the final states a better strategy to estimate multiple observables per iteration may be needed for greater efficiency. One option is using state reconstruction techniques developed in quantum tomography~\cite{Banaszek1999,Christandl2012} or devising strategies tailored to the particular system studied and it's encoding on the quantum computer.

\section{\label{sec:conclusions}Conclusions}

We presented a complete quantum algorithm for calculating the linear response of a quantum system to external perturbations with controllable accuracy. This is achieved by probabilistically preparing the perturbed state (even though a deterministic preparation with polynomial cost is in general available) and then analyzing it by using the standard Phase Estimation Algorithm~\cite{Abrams1999}.
Our approach is efficient (scaling is polynomial in system size and required accuracy) and provides direct access to the final states resulting from the perturbation, a property that potentially makes it extremely valuable to the interpretation of ongoing and planned scattering experiments.

\emph{Acknowledgements:}
We thank G. Perdue for valuable discussions on applications to neutrino scattering from nuclei and M. Savage for discussions and his critical reading of the manuscript.
This research is supported by the the U.S.~Department of Energy, Office of Science, Office of Advanced Scientific Computing Research under contract DE-SC0018223 (SciDAC - NUCLEI) and Office of Nuclear Physics under contract DE-AC52-06NA25396.

%
\end{document}